\documentclass[journal]{IEEEtran}

\usepackage{amsmath,graphicx}
\usepackage{amssymb}
\usepackage{multirow}
\usepackage{booktabs}
\usepackage[T1]{fontenc}
\usepackage{array}
\usepackage{url}
\usepackage{tikz}
\usepackage{tikzscale} %inline display external tikz files

\def\Re{\operatorname{\mathbb{R}e}}
\def\Im{\operatorname{\mathbb{I}m}}

\newcolumntype{?}[1]{!{\vrule width #1}}
\newcommand{\iu}{{i\mkern1mu}}

\title{End-to-End Model for Speech Enhancement by Consistent Spectrogram Masking}
\author{Du Xingjian,
        Zhu Mengyao,
        Shi Xuan,
        Zhang Xinpeng,
        Zhang Wen,
        and Chen Jingdong% <-this % stops a space
\thanks{Du Xingjian, Zhu Mengyao, Shi Xuan and Zhang Xinpeng are with School of Communication and Information, Shanghai University. (Corresponding author: Zhu Mengyao. e-mail: zhumengyao@shu.edu.cn)}% <-this % stops a space
\thanks{Zhang Wen and Chen Jingdong are with Center of Intelligent Acoustics and Immersive Communication, Northwestern Polytechnical University.}% <-this % stops a space
\thanks{Manuscript received 
Sept. 13, 2018. This work was supported by the National Natural Science Foundation of China (61831019) and the Key Support Projects of Shanghai Science and Technology Committee (16010500100). }}
\begin{document}

\maketitle
\begin{abstract}
    Recently, phase processing is attracting increasing interest in speech enhancement community. Some researchers integrate phase estimations module into speech enhancement models by using complex-valued short-time Fourier transform (STFT) spectrogram based training targets, e.g.\ Complex Ratio Mask (cRM) \cite{crmfm}. However, masking on spectrogram would violent its consistency constraints. In this work, we prove that the inconsistent problem enlarges the solution space of the speech enhancement model and causes unintended artifacts. Consistency Spectrogram Masking (CSM) is proposed to estimate the complex spectrogram of a signal with the consistency constraint in a simple but not trivial way. The experiments comparing our CSM based end-to-end model with other methods are conducted to confirm that the CSM accelerate the model training and have significant improvements in speech quality. From our experimental results, we assured that our method could enhance noisy speech audios with both efficiency and effectiveness.
\end{abstract}

\section{Introduction}
\label{sec:intro}
Many of audio and speech processing approaches represent the signal in a time-frequency transformation. The short-time discrete Fourier transform (STFT) are most usually used. After this transformation, the signal can be represented by their magnitude and their phase in complexed value form. However the phase has been largely ignored while the researchers were focusing on the modeling and processing of the STFT magnitude  in the past three decades. \cite{2015Timo}.\par %However, Roux \cite{pcstb} demonstrated that when the inconsistency between magnitude and phase spectrograms is maximized, the same magnitude spectrogram can lead to extremely diverse resynthesized sounds, depending on the phase with which it is combined. Paliwal et al. \cite{tiop} confirmed that enhancing the phase spectrogram of noisy speech leads to perceptual quality improvements. \par

    However, as soon as reconstruction is desired, phase information becomes essential. When the magnitude is modified, it is often sufficient to reuse the original phase to recover the signal, which may lead to undesired artifacts. Some researchers focus on the applications that the original phase is not available \cite{2017Zdenek}. In this case, STFT phase retrieval algorithms construct a new valid phase from the modified magnitude, allowing complete disposal of the existing phase.\par

        %For speech enhancement applications, disposal of the noisy phase may lose the phase information of target speech. 
    Based on phase enhancement research, enhancing the phase spectrogram of noisy speech leads to perceptual quality improvements \cite{tiop}. Instead of separately enhancing the magnitude and phase response of noisy speech, recent researchers focus on jointly enhancing the magnitude and phase responses to further improve the perceptual quality \cite{2016Williamson}. 
    %More specifically, a DNN was employed to estimate the complex ideal ratio mask (cRM) for real and imaginary components.\par
     If the spectrogram is modified, the modified spectrogram may not correspond to the STFT of any time-domain signal anymore, which is so-called inconsistent spectrogram \cite{2015Timo}. The majority of speech enhancement approaches either only modify the magnitude or estimate complex spectrogram, which will most likely lead to an inconsistent spectrogram. It is worth mentioning that consistent spectrogram obtained from the SFTF of a time-domain signal should be a small subset of the complex spectrogram. In this letter, we propose a joint real and imaginary reconstruction algorithm on consistent spectrogram. In other words, given the complex spectrum of noisy speech, we could recover the consistent spectrum of clean speech. Because the optimization space of our method is restricted to a consistent spectrogram, fast convergence rate and high accuracy can be achieved by the proposed speech enhancement algorithm. \par
%    The size of consistent spectrogram should be much less than that of the inconsistent spectrogram. It is apparent that consistent spectrograms can be obtained from a time-domain signal. If STFT and ISTFT are taken into account during the convergence of the neural network, the optimization space is restricted to a consistent spectrogram. Therefore fast convergence rate and high accuracy can be achieved by proposed method.\par
    
This paper is organized as follows. Section \ref{incon} reviews masking based speech enhancement methods and inconsistent spectrogram problem. Section \ref{sec:meth} proposes Consistent Spectrogram Masking algorithm. Section \ref{exp} describes the experimental setups used to evaluate the performance of the model we propose. Finally, Section \ref{conclusion} present conclusions.

\begin{figure*}[t]
    \centering
    \includegraphics[width=1.\textwidth]{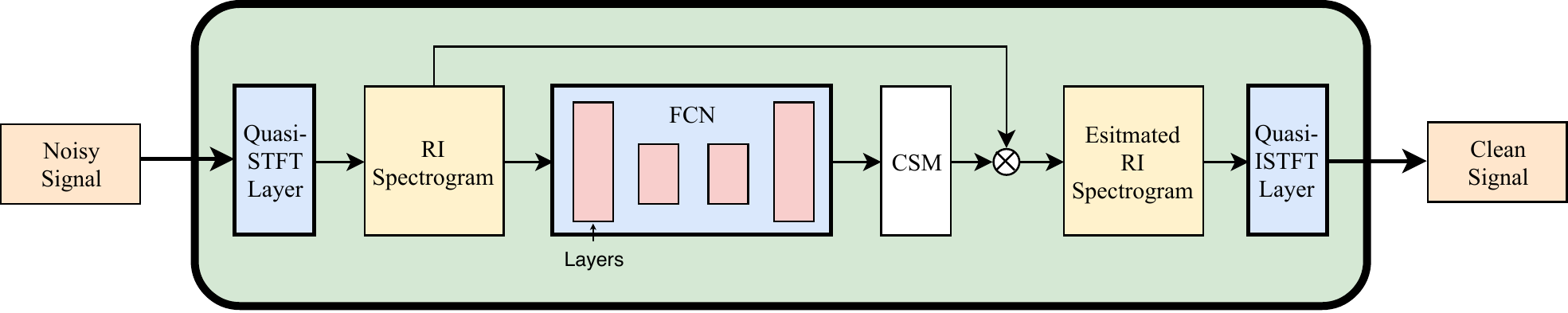}
    \caption{The framework of our proposed end-to-end model for speech enhancement}
    \label{fig:arc}
\end{figure*}

\section{Masking methods and Inconsistent Spectrograms Problem}
\label{incon}
The common speech enhancement setup consisting of STFT analysis, spectral modification, and subsequent inverse STFT (ISTFT). The analyzed digital signal yielding the complex-valued STFT coefficients, this procedure can be compactly described as $S = STFT(x)$. Recently, phase processing has emerged as a further leverage on speech enhancement tasks, including the noticeable work like Phase Sensitive Masking (PSM) \cite{PSM}, and Complex Ratio Masking (cRM) \cite{crmfj,crmfm}. Wang et al. illustrated that the real and imaginary spectrograms exhibits clear temporal and spectral structure, so they propose the cRM which is 
defined as follow:

%Polar coordinates (i.e., magnitude and phase) are commonly used when enhancing the STFT spectrogram of noisy speech, as defined in Eq. \ref{eq:1}.

%\begin{equation}
%    \label{eq:1}
%    \mathit{S}_{t,f} = |\mathit{S}_{t, f}| %e^{i \theta_{\mathit{S}_{t,f}}}
%\end{equation}
%\noindent where $|S_{t,f}|$ represents the magnitude response and $\theta_{\mathit{S}_{t,f}}$ represents the phase response of the STFT spectrogram at time $\mathit{t}$ and frequency $\mathit{f}$. 

 %The common speech enhancement focus on modifying only the magnitude of the STFT components $|\mathit{S}_{t, f}|$, because it was generally considered that most of the insight about the structure of the signal could be obtained from the magnitude, while little information could be obtained from the phase component, such as IRM \cite{irm}, which is defined as follow:

%\begin{equation}
%IRM(t,f) = \left(   {{S^2}_{t,f}} \over  {{S^2}_{t,f} + {N^2}_{t,f}}  \right)^\beta
%\end{equation}
%where $S^2_{t,f}$ and $N^2_{t,f}$ denote the speech and noise energy, respectively, in a particular T-F unit. $\beta$ is a tunable parameter to scale the mask.\par

%%%% emphasize the complex 
\begin{equation}
cRM(t,f) = \frac{\Re\{\mathit{S_{t,f}}\}}{\Re\{\mathit{S_{t,f}+N_{t,f}}\}} + \iu \frac{\Im\{S_{t,f}\}}{\Im\{S_{t,f}+N_{t,f}\}}
\end{equation}

% inconsistant problem
However, the methods mentioned above all ignore the inconsistent spectrogram problem.  
The inconsistent spectrogram problem illustrated by Timo Gerkmon is a great challenge to speech enhancement. Because the STFT analysis is done using overlapping analysis window, any modification for individual signal components (sinusoids, impulses), will be spread over multiple frames and multiple STFT frequencies locations. 

%In other words, neighboring time-frequency points show dependencies due to the STFT analysis for following two reasons. First, because of the finite length of an audio signal, the neighboring frequency bands are not independent. Second, successive frames overlap and hence partly share some information. 

%%%%%%%%%%%%%%%%%%%%%%%%%%%%%%%% Le Roux %%%%%%%%%%%%%%%%%%%%%%%%%%%%%%%%%%
Le Roux et al. \cite{pcstb} derived the consistency constraints for STFT spectrograms consicely. Let $S_{t , f}$ be a set of complex numbers, where $t$ will correspond to the frame index and $f$ to the frequency band index, and $W_a$, $W_s$ are analysis and synthesis window function verifying the perfect reconstruction conditions for a frame shift $R$. For any complex spectrogram $S$, we can get the following equation.\par

\begin{multline*}
STFT(ISTFT(S_{t,f^{\prime}})) 
%= \frac { 1 } { N } \sum _ { k } W_a ( k ) e ^ { - j 2 \pi k \frac { f ^ { \prime } } { N } } \\ 
%\{ W_s ( k ) \sum _ {f = 0 } ^ { N - 1 } S _{ t , f } e ^ { j 2 \pi f \frac{k}{N} } \\
%+ \sum _ { f = 0 } ^ { N - 1 } S_{t - 1, f} e ^ { j 2 \pi f \frac { k +  R } { N } } + \\
% W_s( k - R ) \sum _ { f = 0 } ^ { N - 1 } S_{t + 1 , f} e ^ { j 2 \pi f \frac { k -  R } { N } } \} \\
= S_{t,f} + \frac { 1 } { N } \sum _ {k} W_a ( k ) e ^ { - j 2 \pi k \frac { f ^ { \prime } } { N } } \\
\{W_s( k +  R ) \sum _ { f = 0 } ^ { N - 1 } S_{t - 1, f} e ^ { j 2 \pi n \frac { k + R } { N } } + \\
W_s( k -  R ) \sum _ { f = 0 } ^ { N - 1 } S _{t + 1 , f} e ^ { j 2 \pi n \frac { k - R } { N } } \} \\
\end{multline*}

%%%%%%%%%%%%%%%%%%%%%%%%%%%%%%%%%%%%%%%%%%%%%%%%%%%%%%%%%%%%%%%%%%%%%%%%%%%%
  $S$ can be divided into $S_{con}$ and $S_{incon}$. $S_{con}$ can be obtained from STFT of time signal $x$. And there is a one-to-one mapping between $S_{con}$ and $x$ and a many-to-one mapping between $S_{incon}$ and $x$. The resynthesized time signal ISTFT ($S_{incon}$) has the consistent spectrogram $S_{con}$ after STFT transform. As a consequence, the relation between $S_{con}$ and $S_{incon}$ can be shown in the following equation.
\begin{equation}
\label{scon}
    S_{con} = STFT(ISTFT(S_{incon})) \neq S_{incon}
\end{equation}

%%%%%%%%%%%%%%%%%%%%%%%%%%%%%%%%%%%%%%%%%%%%%%%
%   un-consistent not inversable
%   \hat{S^{\prime}} consistent
%%%%%%%%%%%%%%%%%%%%%%%%%%%%%%%%%%%%%%%%%%%%%%%

\begin{figure}
    \centering
    \includegraphics[width=0.4\textwidth]{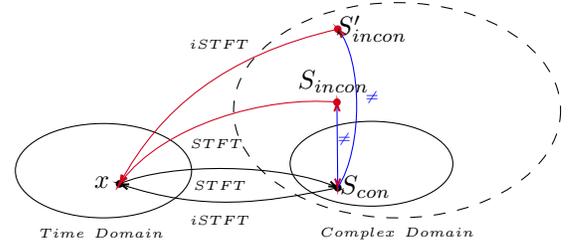}
    \caption{An illustration of the notion of consistency. STFT transform is an injective function which maps distinct valid signals to corresponding consistent spectrograms $S_{con}$ respectively i.e. there is a perfect one-to-one correspondence between the sets of time signal and consistent spectrograms. However, STFT transform is not guaranteed to be invertiable for inconsistent spectrograms $S_{incon}$. There is a many-to-one mapping between $S_{incon}$ and time signal $x$ as indicated by red arrows.}
    \label{inconsist}
\end{figure}

%% jia dao caption
Since the many-to-one mapping between $S_{incon}$ and $x$ and one-to-one mapping between $S_{con}$ and $x$ as illustrated in Fig. \ref{inconsist}, the space of $S_{incon}$ is much larger than the space of $S_{con}$. Therefore, the estimated clean spectrogram $\hat{S}$ in the design of speech enhancement system tend to fall into the inconsistent spectrograms $S_{incon}$ space. The commonly ignored inconsistent spectrograms problem not only introduces artifacts into resynthesized signals because of the inconsistency of overlapping frames but also increases difficulties of model convergence due to the expansion of inconsistent spectrogram space. \par
%which is a difference from the desired spectrogram $\hat{S}$, as illustrated in Figure\ref{fig:inconsist}. Such spectrograms are called inconsistent spectrogram, while consistent spectrograms verify $STFT(ISTFT(S_{con})) = S_{con}$ and can be obtained from a time-domain signal.
%That means the STFT transform is not guaranteed to be invertible for inconsistent spectrograms $S_{incon}$ which is commonly ignored in the design of the speech enhancement system. More specifically, the cost of performance will be incurred by the inconsistent spectrogram problem, a) The inconsistency of overlapping frames will introduce artifacts into time-domain signals. b) More specific to deep learning based speech enhancement methods, increasing difficulties are faced by the convergence of model due to the expansion of searching space while the desired target is a consistent spectrogram. The set of the consistent spectrogram is belonged to the universe of all spectrograms, as reconfirmed by the t-SNE\cite{tsne} visualization of spectorgrams\ref{distribution}.

%\begin{figure}
%    \centering
%    \includegraphics[width=0.4%\textwidth]{tsne.pdf}
%    \caption{t-SNE\cite{tsne} visualization of spectrograms from randomly generated signals, For visualization purposes, embeddings are colored with red and blue, the blue points represent consistent spectorgram, the red points represent their inconsistent version.}
%    \label{distribution}
%\end{figure}

%Therefore, we introduce a Quasi-layer(QL) which integrates Fourier transform into our deep learning framework. 
%Because the Quasi-Layer can be trained with backpropagation, we propose to jointly estimate the real and imaginary mask using an objective function defined in the time domain. 
%As the loss is calculated in the time domain; our method avoids inconsistent spectrogram problem.
%And the single objective function makes our model easier to converge than the multi-objective optimization\cite{csebc}. 
%Different from research tried to build end-to-end models on speech waveforms directly\cite{etewu}, our model with Quasi-layer predict Real and Imaginary mask in complex spectrogram.\par

\section{Consistent Spectrogram Masking}
\label{sec:meth}
\subsection{Masking with Consistency constraints}
The most of model-based speech enhancement methods can be regarded as minimize the follow objective function:
\begin{equation}
\label{ori}
%\textit{O} = (\hat{S} - STFT(x))^\beta
\textit{O} = ||\hat{S}- STFT(x)||^\beta \\
\end{equation}
where $\hat{S}$ is estimated clean spectrogram, $x$ denotes clean signal i.e the ground truth for the model, and $\beta$ is a tunable parameter to scale the distance.\par 
Because $\hat{S}$ is estimated from a non-linear function of nosiy speech $F(S+N)$ (non-linear function can be neural network or HMM etc.), these non-linear operation may destruct the corresponding relationship between neirbouring frames and can not guarantee the consistence of $\hat{S}$. As a result, the objective function defined in spectrogram incurs the aforementioned inconsistent spectrogram problem. Here we derive the difference between objective functions defined in consistent and inconsistent spectrogram. \par

If we apply both ISTFT and STFT transform in terms of Eq. \ref{ori}, we can have the following equations. Since the consistency of $\hat{S}$ that the model estimate cannot be guaranteed, $\hat{S}_{con} = STFT(ISTFT(\hat{S}))$ can be deduced from Eq. \ref{scon} and $\hat{S}_{con}$ is not equal to $\hat{S}$. Therefore, the following objective functions are not equal to the objective function in Eq. \ref{ori}. It worth noting that the last two equations in Eq. \ref{noob} shows the equivalent form of objective functions on both time domain and consistent spectrogram.

\begin{equation}
\label{noob}
\begin{aligned}
\centering
%\textit{O} = &||\hat{S}- STFT(x)||^\beta \\
%\not\equiv & ||iSTFT(\hat{S}) - iSTFT(STFT(x))||^\beta \\
&||STFT(ISTFT(\hat{S})) - STFT(ISTFT(STFT(x)))||^\beta \\
 = &||\hat{S}_{con} - STFT(x)||^\beta  \\
%\equiv  &||iSTFT(\hat{S}_{con}) - iSTFT(STFT(x))||^\beta \\
 = &||ISTFT(\hat{S}_{con}) - x||^\beta 
\end{aligned}
\end{equation}
% O_con is consistent objective loss

%We consider, as noted in section \ref{incon}, the optimization target of model set on spectrogram domain and the paradigm that output masked or modified spectrograms directly would introduce artifacts during the phase of ISTFT transform and enlarge the solution space while overlooking the consistency spectrogram of the spectrogram.\par
Follow the motivations noted in section \ref{incon} and the derivation of Eq \ref{noob}, we naturally considered introducing a objective function termed $\textit{O}_{con}$ which is defined on consistent spectrogram domain $\hat{S}_{con}$. We name our method as Consistent Spectrogram Masking (CSM) because it iteratively minimizes the objective function and derives masking on a consistent spectrogram. Our proposed method could dispel the artifacts of resynthesis signal and speed up of model training based on space contraction on a consistent spectrogram. \par

\begin{equation}
\label{of}
\begin{aligned}
\centering
%    O_{con} = &||\hat{S}_{con} - STFT(x)||^\beta \\
%      = &||STFT(iSTFT(\hat{S})) - STFT(x)||^\beta
     O_{con} = &||ISTFT(\hat{S}) - x||^\beta
\end{aligned}
\end{equation}

Although $\hat{S}_{con}$ and $\hat{S}$ are different, $ISTFT(\hat{S}_{con})$ and $ISTFT(\hat{S})$ are the same in time domain (illustrated by Fig. \ref{inconsist} and Eq. \ref{scon}). Thus, we have the useful form of objective function in Eq. \ref{of}. By coincidence, there are some similarities between the Eq. \ref{of} and Griffin-Lim algorithm \cite{griffin}, because a lot of ISTFT and STFT calculations are needed in the optimization procedure. In Griffin-Lim algorithm, phase information is solely derived from the magnitude of the spectrogram. 
Nevertheless, our method could estimate both magnitude and phase information in the form of complex numbers on the consistent spectrogram.
Thus, we defined Consistent Spectrogram Masking (CSM) as follow by given the complex spectrogram of noisy speech,$Y_{t,f}$
\begin{equation}
\begin{aligned}
    \centering
    \hat{S}_{t,f} = M\hspace{-3pt}R_{t,f}  \Re\{Y_{t,f}\} + \iu \hspace{1pt} M\hspace{-2.4pt}I_{t,f}  \Im\{Y_{t,f}\} 
\end{aligned}
\end{equation}
where $M\hspace{-3pt}R_{t,f}$, $M \hspace{-2.4pt}I_{t,f}$ represent the mask for the real and imaginary spectrogram at time $t$ and frequency $f$. \par

%To keep the consistency constraints for spectrograms, we propose a simple but not trivial methods: the end-to-end model output the estimated clean signal in time domain by a differential module termed Quasi-Layers, which simulates the STFT operation. So we can define the objective function in the time domain.

%%% bu yao QL
%%%% consistent pros
%Performing loss calculation in the time domain can avoid inconsistent audio signals introduced by other methods based on frequency-domain loss functions. The reasoning behind this method is simple, as we calculate training in time domain directly, we are approaching the output audio with the clean signal, the inconsistent spectrogram problem is circumscribed as we get a valid audio signal, whose spectrogram is, of course, consistent, approaching the desired output directly.\par

%gL algorithm

\subsection{The framework of our proposed end-to-end model}
\label{ssec:ql}

%%%%%%%%%%%%%%%%%%%%%%%%%%%%%%%%%%%%%%%%%%%%%%%%%%%%%%%%%%%%%%
%%%%%%%%%%%%%%%%%%%%%%% To implemented %%%%%%%%%%%%%%%%%%%%%%%
%%%%%%%%%%%%%%%%%%%%%%%%%%%%%%%%%%%%%%%%%%%%%%%%%%%%%%%%%%%%%%
Following the aforementioned methodology and principle that optimizing the model with consistency constraint, we designed an end-to-end speech enhancement model which comprises a densely connected convolutional neural network (CNN) and integrated Quasi-Layers (QL). A high-level visual depiction of our proposed model is presented in Fig. \ref{fig:arc}. Specifically, for corresponding functionalities, the CNN module is employed to adaptively modify spectrogram of the input signal, and QL is a backpropagate module designed to simulate the STFT transform and its inversion, thereby making it possible to directly accumulate the loss on consistent spectrogram. \par
%In the ensuing discussion, we would explain these two-part precisely.
The CNN based acoustic models have been used in speech enhancement and source separation tasks and have been proven to improve the performance \cite{etewu}. The unique connection structure and weight sharing make CNN capable of learning feature representation via applying convolutional filters to the spectrogram of audio. However, there is an intrinsic tradeoff problem between kernel size and feature resolution.
In other words, a larger kernel can exploit more contextual information in time dimension or learning pattern in a wider band, but obtain lower resolution features.
In this work, we utilize a densely connected fully convolutional network (FCN) \cite{densenet} which can learn multi-scale features efficiently to solve the trade-off problem. 
In a standard feedforward network, the output of the $l$th layer is computed as $x_l = H_l(x_{l-1}),$ where the network input is denoted as $x_{l-1}$ and $H_l(\cdot)$ is a nonlinear transformation which can be a composite function of operations such as nonlinear activation, pooling or convolution\cite{densenet}. \par
The idea of DenseNet is to use concatenation of feature maps produced in preceding layers as the input to succeeding layers:
\begin{equation}
    x_l = H_l([x_{l-1}, x_{l-2},\ldots,x_0]),
    \label{eqDenseNet}
\end{equation}
where $[x_{l-1}, x_{l-2},\ldots,x_0]$ refers to the concatenation of the feature maps produced in layers $0, \ldots, l-1$ \cite{densenet}. Such dense connectivity enables all layers not only to receive the gradient directly but also to reuse features computed in preceding layers. This pipeline avoids the re-calculation of similar features in different layers and makes network can learn different level features in the same layer \cite{densenet}. The experimental results show that our DenseNet based approach has a considerable improvement compared to DNN based model. \par

The FCN is the backbone of our model, and the preprocessing and postprocessing modules Quasi-Layers, are also vital parts of the whole system. The Quasi-STFT layer uses two 1-dimensional convolutions, each of which is initialized with real and imaginary part of discrete Fourier transform kernels respectively, following the definition of STFT: 
\begin{equation}
    S_{t,f} = \sum_{n=0}^{N-1}x_{Nt +n}\cdot[\cos(2\pi f n/N)-i\cdot \sin(2\pi fn/N)] 
\end{equation}
\noindent for $k\in~[0,N-1]$,
the Quasi-ISTFT layer is similar to this one. These modules are constructed on normal convlutional layers and thus it's easily to integrate these modules into the neural network based model. 
These Quasi-Layers can bring us benefits in two folds, firstly Quasi-ISTFT also offers the probability to define the objective function on a consistent spectrogram as Eq. \ref{of}. On the other hand, the integration of STFT and ISTFT into the end-to-end model can make Fourier transform kernel and window function learnable with the backpropagation. \par

%Overall, we have presents the main insights with this CSM based end-to-end model. We conceptualize the idea of estimating the clean signal on consistent spectrogram domain $S_{con}$ within the studying of the consistency constraints for STFT spectrograms. Meanwhile, we choose to mask on complex spectrogram of the noisy signal, seeking to exploit the information embedded in real and imaginary spectrograms and estimate them jointly rather than processing magnitude spectrograms solely, regarding the importance of phase processing. \par   

%To enhance a noisy speech signal with this model, the Quasi-STFT layer extracts the real and imaginary (RI) spectrogram of the signal firstly, then feeds it into fully convolutional neural network (FCN) to estimate CSM, applies it to noise RI spectrogram to retrieve estimated clean RI spectrogram and finally the Quasi-ISTFT layer output clean signal.
%The Experiment section \ref{exp} gives the promising results of our proposed algorithm.                                                                                                                                                                                                                                                                                                                                                    
\section{Experiment}
\label{exp}
\subsection{Experimental Setup}
\label{ssec:setup}

We conducted our experiments on the Center for Speech Technology Voice Cloning Toolkit (VCTK) \cite{vctk} and The DARPA TIMIT Acoustic-Phonetic Continuous Speech Corpus (TIMIT) \cite{timit} corpora, the training data is supplied by VCTK which includes 400 x 109 sentence uttered by 109 native speakers of English with various accents and the model is evaluated in TIMIT. 
Training and testing in different dataset promise the reliability of results. 
Moreover, the following broadband noise:  speech babble (Babble), cafeteria (Cafe),  factory floor noise (Factory), transportation noise (Road). 
The training set is composed by combining ten random parts from the first half of each noise with each training sample at different SNR levels which is -6, -3, 0, 3 and 6 $dB$ respectively. 
The test set is generated by mixing 60 clean utterances of the last half of the above noises at different SNRs. Dividing noises into two halves ensures that the testing noise segments are unseen during training.\par
The proposed model termed QL-FCN-CSM is given in Figure. \ref{fig:arc}.
Ahead of the FCN, the raw audio input of 66048 samples, is transformed to a 512 x 16 x 2 matrix by STFT Quasi-layer, the window length and hop length of which are set to 1024 and 512 respectively. 
Mean, and variance normalization was applied to the input vector to make the training and testing process stable.  
The perceptual evaluation of speech quality (PESQ) \cite{pesq} and the signal to noise ratio (SNR)  are used to evaluate the quality and intelligibility of different signals. \par

\subsection{Experimental Results}
\subsubsection{Comparison Between Different Objective Functions}
We conducted the experiments with models based on different objective functions, 
the model which is targeted to minimize the error between the complex spectrogram of clean speech and its noisy version is denoted as QL-FCN-cRM (similar to QL-FCN-CSM, but replace CSM with cRM), and the model which estimate magnitude solely is denoted as QL-FCN-IRM (still similar to QL-FCN-CSM, but replace CSM with IRM). \par
Table 1 shows that there is a substantial performance gap between QL-FCN-CSM and QL-FCN-cRM, between QL-FCN-CSM and QL-FCN-IRM, 
which proves the efficiency of CSM which optimize model with the objective function defined in the consistent spectrogram and synthesize waveforms directly.
It is observed that the average PESQ scores and SNR of QL-FCN-CSM and QL-FCN-cRM are always better than the other models,
which proves the effectiveness of the end-to-end model we proposed. Our best results on 0dB condition are even more encouraging: the PESQ score is 0.38 higher than the DNN-cRM, which is state-of-the-art DNN approach.\par
\begin{figure}
    \centering
    \includegraphics[width=0.45\textwidth]{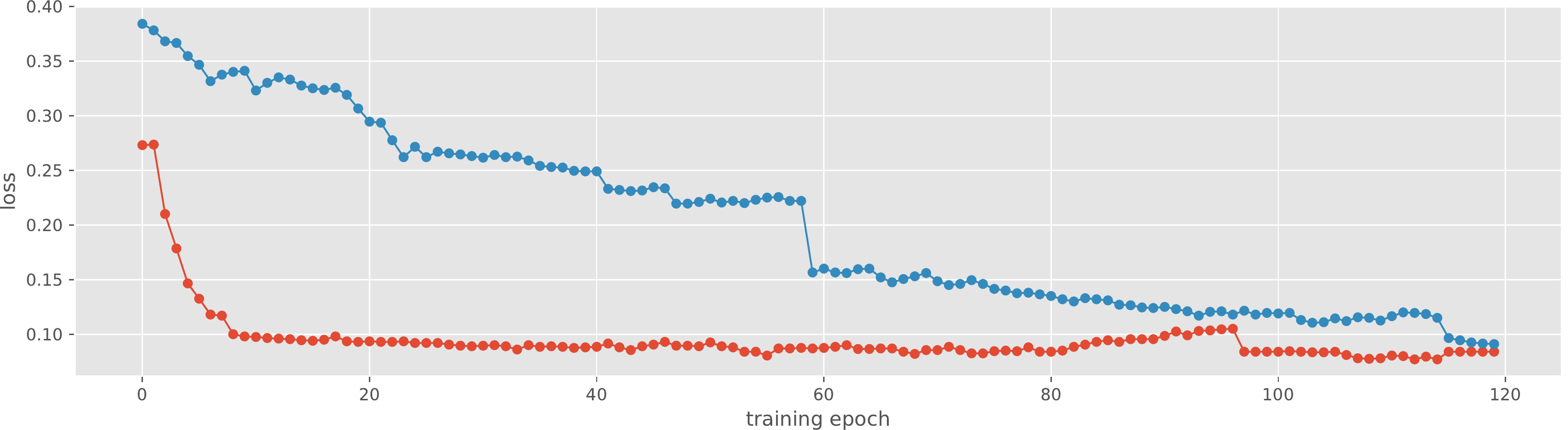}
    \caption{Training CSM-QL and cRM model on VCTK dataset. The preformance of CSM-QL surpass the cRM model with the faster convergence speed.}
    \label{curve}
\end{figure}

It was noteworthy that the convergency speed of QL-FCN-CSM overtaking the others with better performance, these circumstances reinforce the view we hold: the constrain of the estimated spectrogram into the scope of the consistent spectrogram, leading the faster convergence shown in Fig. \ref{curve}.
\subsubsection{Comparison Between Different Network Architectures}
To compare our FCN based model with those base on DNN, experiments compare ours with DNN-cRM \cite{crmfm} (QL is not conducted as there is no convolution procedure here, deep neural network is used instead of FCN) and DNN-IRM \cite{sebodnnwsc}. \par
From Table 1, we can observe that QL-FCN-CSM and QL-FCN-cRM outperform DNN-cRM and DNN-IRM all the time. The results proved the efficiency of our selection of network architecture.
However, the results of QL-FCN-CSM is comparable to those of QL-FCN-cRM in 6 $dB$ and -6 $dB$ conditions. 
It is because artifacts caused by the loss of phase information are negligible in very high or very low SNR conditions \cite{setp}. \par
%It demonstrates that Quasi-Layers brings significant improvement in SNR metric because the objective function defined in the consistent spectrogram can reduce the artificial noise of resynthesis signal. 
%It is also interesting to mention that FCN-CSM outperform DNN-cRM, which illustrate the good feature representation of FCN.

\begin{table}[ht]
  \label{tab:test1}
  \caption{PESQ and SNR performance for the 5 models: No enhancement (a), QL-FCN-CSM (b), QL-FCN-cRM (c), QL-FCN-IRM (d), DNN-cRM (e), DNN-IRM (f).}
  \centering
  \resizebox{0.48\textwidth}{!}{
    \begin{tabular}{ l | c | c c c c c ?{1.5pt} c c c c c  }
    \toprule
     &  & \multicolumn{5}{c?{1.5pt}}{PESQ}  & \multicolumn{5}{c}{SNR} \\
    \toprule
         &SNR  & -6 & -3 & 0 & 3 & 6 & -6 & -3 & 0 & 3 & 6\\
    \midrule 
    \multirow{ 6}{*}{\rotatebox[origin=c]{90}{Babble}} 
       & a & 1.179     & 1.301 & 1.489 & 1.672 & 1.998 & -6.00   &  -3.00  &  0.00 &  3.00   &  6.00  \\
        & b & 1.951     & \bf{2.112} & \bf{2.682} & \bf{2.855} & \bf{3.106} & \bf{5.93} & \bf{8.47} & \bf{11.32} & 13.82 & \bf{16.43} \\
        & c & 1.953     & 2.068 & 2.543 & 2.833 & 2.966 & 5.89 & 8.13 & 10.76 & \bf{13.91} & 16.14\\
        & d & \bf{1.967}& 2.077  & 2.515 & 2.710 & 2.976 & 5.92 & 8.07 & 10.83 & 13.16 &  15.66\\
        & e & 1.914     & 1.836  & 2.299 & 2.517 & 2.843 & 4.67 & 6.87 & 8.38 & 10.98 & 14.73 \\
        & f & 1.809     & 1.787  & 2.113 & 2.442 & 2.798 & 4.09 & 6.53 & 8.05 & 10.12 & 13.09 \\
    \midrule
      
    \multirow{ 6}{*}{\rotatebox[origin=c]{90}{Cafe}} 
     & a & 1.413 & 1.676 & 1.894 & 2.123 & 2.342 & -6.00 & -3.00 & 0.00 & 3.00 & 6.00 \\
    & b & \bf{2.365} & \bf{2.517} & \bf{2.720} & 2.878 & \bf{3.021} & \bf{6.34} & \bf{8.59} & \bf{11.42} & 14.0 & \bf{16.47} \\
    & c & 2.363 & 2.501 & 2.686 & \bf{2.880} & 3.004 & 6.30 & 8.37 & 10.96 & \bf{14.03} & 16.18 \\
    & d & 2.362 & 2.496 & 2.690 & 2.836 & 2.975 & 6.29 & 8.28 & 11.01 & 13.26 & 15.7 \\
    & e & 2.272 & 2.426 & 2.516 & 2.698 & 2.937 & 5.01 & 7.23 & 8.58 & 11.12 & 15.03 \\
    & f & 2.240 & 2.401 & 2.493 & 2.647 & 2.833 & 4.59 & 6.85 & 8.24 & 10.44 & 13.22 \\
     \midrule
  
    \multirow{ 6}{*}{\rotatebox[origin=c]{90}{Factory}} 
    & a & 0.987 & 1.119 & 1.265 & 1.468 & 1.695 & -6.00 & -3.00 & 0.00 & 3.00 & 6.00 \\
    & b & \bf{1.783} & \bf{1.911} & \bf{2.121} & \bf{2.304} & \bf{2.460} & \bf{7.16} & \bf{8.82} & \bf{11.58} & \bf{14.19} & \bf{16.53} \\
    & c & 1.778 & 1.89 & 2.106 & 2.302 & 2.441 & 7.10 & 8.55 & 11.37 & 14.16 & 16.25 \\
    & d & 1.78 & 1.893 & 2.101 & 2.246 & 2.408 & 7.12 & 8.59 & 11.30 & 13.36 & 15.75 \\
    & e & 1.687 & 1.813 & 1.908 & 2.113 & 2.381 & 5.89 & 7.55 & 8.78 & 11.47 & 15.33 \\
    & f & 1.625 & 1.765 & 1.874 & 2.046 & 2.240 & 5.09 & 6.93 & 8.34 & 10.55 & 13.27 \\
    \midrule
      
    \multirow{ 6}{*}{\rotatebox[origin=c]{90}{Road}} 
    & a & 2.182 & 2.363 & 2.547 & 2.721 & 2.903 & -6.00 & -3.00 & 0.00 & 3.00 & 6.00 \\
    & b & \bf{2.995} & \bf{3.095} & \bf{3.265} & \bf{3.405} & 3.529 & \bf{7.46} & \bf{9.03} & \bf{11.74} & \bf{14.28} & 16.63 \\
    & c & 2.982 & 3.084 & 3.253 & 3.403 & \bf{3.530} & 7.26 & 8.88 & 11.53 & 14.25 & \bf{16.65} \\
    & d & 2.98 & 3.078 & 3.249 & 3.356 & 3.493 & 7.22 & 8.79 & 11.45 & 13.43 & 15.89 \\
    & e & 2.905 & 3.007 & 3.084 & 3.253 & 3.467 & 6.03 & 7.64 & 8.87 & 11.53 & 15.39 \\
    & f & 2.853 & 2.966 & 3.059 & 3.185 & 3.352 & 5.19 & 7.01 & 8.47 & 10.42 & 13.37 \\
    \bottomrule
  \end{tabular}}

\end{table}

\begin{figure}[ht]
    \centering
    \includegraphics[width=0.48\textwidth]{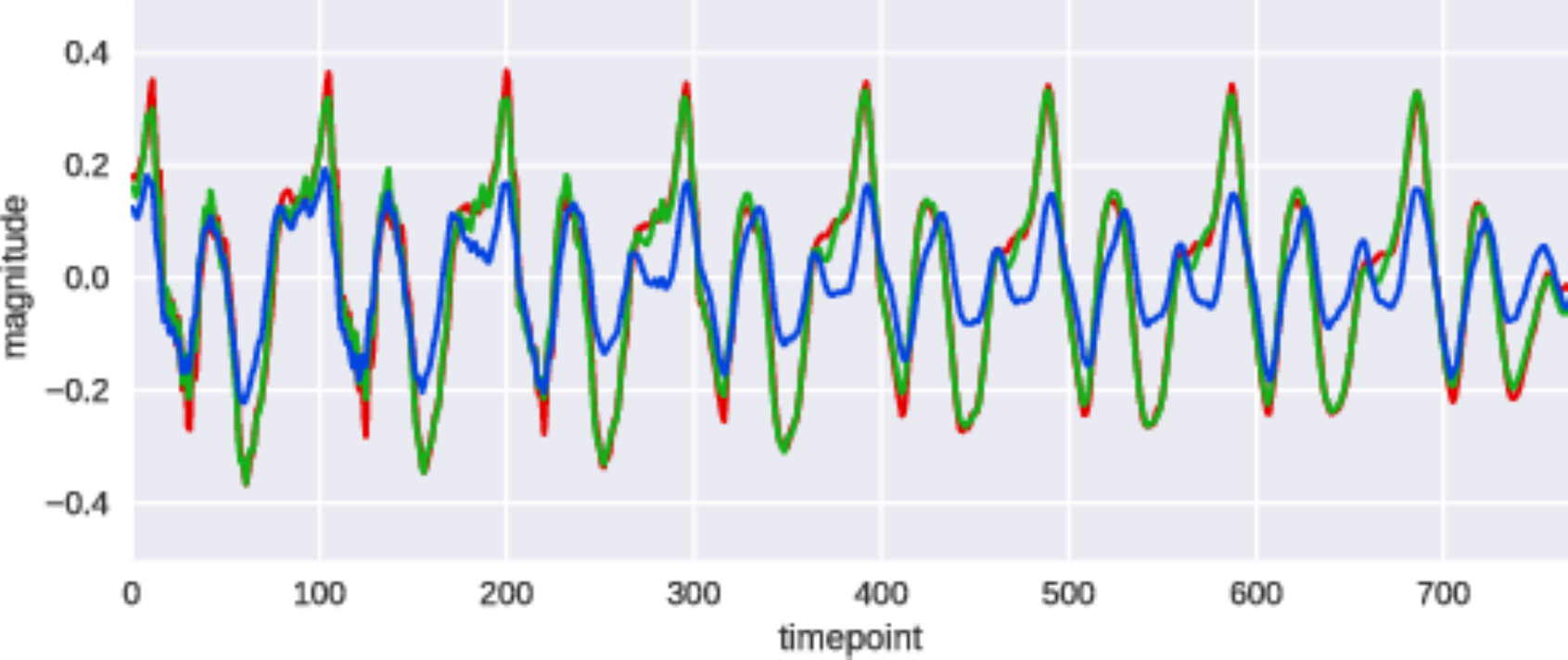}

    \caption{A random clip (768 samples) from the waveform of the experimental results. Red line indicates the clean signal. The green line and the red line indicate the output of QL-FCN-CSM and QL-FCN-IRM respectively. It is obvious that estimating spectrogram masks in a consistent manner can reduce distortion of results in the time domain.}

    \label{fig:wave}
\end{figure}

% dierduan

\section{Conclusions}
\label{conclusion}
The insights and deductions of our work are clear and comprehensive. We draw concepts from prior works that a) Phase processing is essential to speech enhancement tasks; b)  Masking on spectrogram would destruct the consistency constraints.
In this letter, we unveil facts that inconsistent spectrograms problem slow the convergence of model and cause unintended artifacts. To estimate the clean spectrogram (including magnitude and phase) from the STFT of noisy speech with the constraint of consistency, we design a CSM on complex spectrogram and derive the loss function in the consistent spectrogram, which resolves the problem of inconsistent spectrogram and phase processing simultaneous and jointly.
%Because of methods, e.g., spectral subtraction, only modifying magnitude then synthesizing phase through GL algorithm or a model with loss calculated among complex spectrogram, cannot meet two conditions as mentioned earlier concurrently. 

In technical details, we implement new Quasi-Layers to emulate STFT with convolution layers in the neural network, which makes it possible to optimize our model with an objective function on the consistent spectrogram.
%Besides, this implementation accelerates the computation of model and make it possible to fine-tune STFT kernels in a neural network, which we will discuss in more details in our future work. 
DenseNet is selected as the basis of our model framework rather than vanilla CNN or DNN, for its superior ability to extract features with various scales in a spectrogram. The experimental results show that the considered acceleration of convergence and the improvement of quality occurred.  
\par

%The contribution of this paper is three-fold.
%First, we define a novel joint real and imaginary mask on the complex spectrogram, it can be estimated by deep learning based model as the experiments indicate.
%Second, we design Quasi-Layers, which is mathematically equal to STFT and ISTFT, to capacitate model to process waveform input and generate output directly in a waveform. 
% This module can be integrated into existing learning based acoustic model easily. 
%Last but not least, we propose an end-to-end model combined the Quasi-Layer, DensetNet like fully convolution neural network and JRIM to tackle speech enhancement problem, and the results outperform the state-of-the-art neural network-based model. We believe our work will shed lights on the future deep learning based acoustic model. 
%In the future, we will investigate the application of our model in speech separation and speech dereverberation.

\bibliographystyle{IEEEtran}

\bibliography{template.bib}

\begin{thebibliography}{10}
\providecommand{\url}[1]{#1}
\csname url@samestyle\endcsname
\providecommand{\newblock}{\relax}
\providecommand{\bibinfo}[2]{#2}
\providecommand{\BIBentrySTDinterwordspacing}{\spaceskip=0pt\relax}
\providecommand{\BIBentryALTinterwordstretchfactor}{4}
\providecommand{\BIBentryALTinterwordspacing}{\spaceskip=\fontdimen2\font plus
\BIBentryALTinterwordstretchfactor\fontdimen3\font minus
  \fontdimen4\font\relax}
\providecommand{\BIBforeignlanguage}[2]{{%
\expandafter\ifx\csname l@#1\endcsname\relax
\typeout{** WARNING: IEEEtran.bst: No hyphenation pattern has been}%
\typeout{** loaded for the language `#1'. Using the pattern for}%
\typeout{** the default language instead.}%
\else
\language=\csname l@#1\endcsname
\fi
#2}}
\providecommand{\BIBdecl}{\relax}
\BIBdecl

\bibitem{crmfm}
D.~S. Williamson, Y.~Wang, and D.~Wang, ``Complex ratio masking for monaural
  speech separation,'' \emph{{IEEE/ACM} Trans. Audio, Speech {\&} Language
  Processing}, vol.~24, no.~3, pp. 483--492, 2016.

\bibitem{2015Timo}
\BIBentryALTinterwordspacing
T.~Gerkmann, M.~Krawczyk{-}Becker, and J.~L. Roux, ``Phase processing for
  single-channel speech enhancement: History and recent advances,''
  \emph{{IEEE} Signal Process. Mag.}, vol.~32, no.~2, pp. 55--66, 2015.
  [Online]. Available: \url{https://doi.org/10.1109/MSP.2014.2369251}
\BIBentrySTDinterwordspacing

\bibitem{2017Zdenek}
\BIBentryALTinterwordspacing
Z.~Prusa and P.~Rajmic, ``Toward high-quality real-time signal reconstruction
  from {STFT} magnitude,'' \emph{{IEEE} Signal Process. Lett.}, vol.~24, no.~6,
  pp. 892--896, 2017. [Online]. Available:
  \url{https://doi.org/10.1109/LSP.2017.2696970}
\BIBentrySTDinterwordspacing

\bibitem{tiop}
K.~K. Paliwal, K.~K. W{\'{o}}jcicki, and B.~J. Shannon, ``The importance of
  phase in speech enhancement,'' \emph{Speech Communication}, vol.~53, no.~4,
  pp. 465--494, 2011.

\bibitem{2016Williamson}
\BIBentryALTinterwordspacing
D.~S. Williamson, Y.~Wang, and D.~Wang, ``Complex ratio masking for joint
  enhancement of magnitude and phase,'' in \emph{2016 {IEEE} International
  Conference on Acoustics, Speech and Signal Processing, {ICASSP} 2016,
  Shanghai, China, March 20-25, 2016}.\hskip 1em plus 0.5em minus 0.4em\relax
  {IEEE}, 2016, pp. 5220--5224. [Online]. Available:
  \url{https://doi.org/10.1109/ICASSP.2016.7472673}
\BIBentrySTDinterwordspacing

\bibitem{PSM}
\BIBentryALTinterwordspacing
H.~Erdogan, J.~R. Hershey, S.~Watanabe, and J.~L. Roux, ``Phase-sensitive and
  recognition-boosted speech separation using deep recurrent neural networks,''
  in \emph{2015 {IEEE} International Conference on Acoustics, Speech and Signal
  Processing, {ICASSP} 2015, South Brisbane, Queensland, Australia, April
  19-24, 2015}, 2015, pp. 708--712. [Online]. Available:
  \url{https://doi.org/10.1109/ICASSP.2015.7178061}
\BIBentrySTDinterwordspacing

\bibitem{crmfj}
D.~S. Williamson, Y.~Wang, and D.~Wang, ``Complex ratio masking for joint
  enhancement of magnitude and phase,'' in \emph{2016 {IEEE} International
  Conference on Acoustics, Speech and Signal Processing, {ICASSP} 2016,
  Shanghai, China, March 20-25, 2016}.\hskip 1em plus 0.5em minus 0.4em\relax
  {IEEE}, 2016, pp. 5220--5224.

\bibitem{pcstb}
J.~{Le Roux}, ``Phase-controlled sound transfer based on maximally-inconsistent
  spectrograms,'' in \emph{Proceedings of the Acoustical Society of Japan
  Spring Meeting}, no. 1-Q-51, Mar. 2011.

\bibitem{griffin}
S.~Nawab, T.~Quatieri, and J.~Lim, ``Signal reconstruction from short-time
  fourier transform magnitude,'' \emph{IEEE Transactions on Acoustics, Speech,
  and Signal Processing}, vol.~31, no.~4, pp. 986--998, 1983.

\bibitem{etewu}
S.~Fu, Y.~Tsao, X.~Lu, and H.~Kawai, ``End-to-end waveform utterance
  enhancement for direct evaluation metrics optimization by fully convolutional
  neural networks,'' \emph{CoRR}, vol. abs/1709.03658, 2017.

\bibitem{densenet}
G.~Huang, Z.~Liu, L.~van~der Maaten, and K.~Q. Weinberger, ``Densely connected
  convolutional networks,'' in \emph{Proceedings of the IEEE Conference on
  Computer Vision and Pattern Recognition}, 2017.

\bibitem{vctk}
C.~Veaux, J.~Yamagishi, K.~MacDonald \emph{et~al.}, ``Cstr vctk corpus: English
  multi-speaker corpus for cstr voice cloning toolkit,'' 2017.

\bibitem{timit}
J.~S. Garofolo, L.~F. Lamel, W.~M. Fisher, J.~G. Fiscus, and D.~S. Pallett,
  ``Darpa timit acoustic-phonetic continous speech corpus cd-rom. nist speech
  disc 1-1.1,'' \emph{NASA STI/Recon technical report n}, vol.~93, 1993.

\bibitem{pesq}
A.~W. Rix, J.~G. Beerends, M.~P. Hollier, and A.~P. Hekstra, ``Perceptual
  evaluation of speech quality (pesq)-a new method for speech quality
  assessment of telephone networks and codecs,'' in \emph{{IEEE} International
  Conference on Acoustics, Speech, and Signal Processing, {ICASSP} 2001, 7-11
  May, 2001, Salt Palace Convention Center, Salt Lake City, Utah, USA,
  Proceedings}.\hskip 1em plus 0.5em minus 0.4em\relax {IEEE}, 2001, pp.
  749--752.

\bibitem{sebodnnwsc}
M.~Tu and X.~Zhang, ``Speech enhancement based on deep neural networks with
  skip connections,'' in \emph{2017 {IEEE} International Conference on
  Acoustics, Speech and Signal Processing, {ICASSP} 2017, New Orleans, LA, USA,
  March 5-9, 2017}.\hskip 1em plus 0.5em minus 0.4em\relax {IEEE}, 2017, pp.
  5565--5569.

\bibitem{setp}
P.~C. Loizou, \emph{Speech enhancement: theory and practice}.\hskip 1em plus
  0.5em minus 0.4em\relax CRC press, 2013.

\end{thebibliography}

\end{document}